# Conformal Killing Vectors Of Plane Symmetric Four Dimensional Lorentzian Manifolds

Suhail Khan[*1], Tahir Hussain[1], Ashfaque H. Bokhari[2] and Gulzar Ali Khan[1]

[1]*Department of Mathematics, University of Peshawar, Peshawar Khyber Pakhtoonkhwa, Pakistan.*

[2]*Department of Mathematics, King Fahd University of Petroleum and Minerals, Dhahran 31261, Saudi Arabia.*

[*]*Email: suhail_74pk@yahoo.com*

**ABSTRACT**

In this paper, we investigate conformal Killing's vectors (CKVs) admitted by some plane symmetric spacetimes. Ten conformal Killing's equations and their general forms of CKVs are derived along with their conformal factor. The existence of conformal Killing's symmetry imposes restrictions on the metric functions. The conditions imposing restrictions on these metric functions are obtained as a set of integrability conditions. Considering the cases of time-like and inheriting CKVs, we obtain spacetimes admitting plane conformal symmetry. Integrability conditions are solved completely for some known non-conformally flat and conformally flat classes of plane symmetric spacetimes. A special vacuum plane symmetric spacetime is obtained, and it is shown that for such a metric CKVs are just the homothetic vectors (HVs). Among all the examples considered, there exists only one case with a six dimensional algebra of special CKVs admitting one proper CKV. In all other examples of non-conformally flat metrics, no proper CKV is found and CKVs are either HVs or Killing's vectors (KVs). In each of the three cases of conformally flat metrics, a fifteen dimensional algebra of CKVs is obtained of which eight are proper CKVs.

## 1. INTRODUCTION

The general theory of Relativity is governed by the highly non-linear Einstein's Field Equations (EFEs). Due to this non-linearity, it is quite difficult to find exact solutions of EFEs. A list of physically interesting exact solutions of EFEs is documented in [1]. In order to understand an existing link between the structure of spacetime and the gravitational interaction, exact solutions of EFEs can be classified according to different symmetries possessed by the spacetime metrics representing them. In general Relativity, these spacetime symmetries are of interest because their existence is directly related to the presence of conservation laws [2]. Apart from general Relativity, existence of conservation laws are also of pivotal interest for all physical systems which are expressed in terms of the invariance property of physical systems under a continuous symmetry. In particular, a physical system admits energy conservation law if it is invariant under time translation. An analogue of energy conservation law in General Relativity is defined by the invariance property of spacetime metric under a time translation. In Relativity conservation laws determine physical characteristics possessed by the solutions of the Einstein filed equations, which are symmetries represented by vector fields associated with local diffeomorphisms [3]. These symmetries represented by vector fields are associated with local diffeomorphisms that characterize certain types of geometrical structures [3]. A KV is an example of such symmetry and it preserves the spacetime metric tensor. On the other hand, it is not always possible to find conservation laws via KVs. In such cases, conformal transformations are employed that may provide conservation laws not given by KVs. Geometrically, these conformal transformations leave the light cone structure and Maxwell's law of electromagnetic theory invariant. Conservation laws provided by conformal transformations are interpreted in terms of existence of conformal conservation laws or conformal KVs. An example of conformal KV are the famous Friedman metrics for which where there does not exist a time translational invariance giving well defined energy conservation law, it does admit a conformal time translational invariance providing a conformal analogue of energy conservation. A set of well-known symmetries may include HVs (along which the metric tensor is preserved up to a constant scale factor), CKVs (along which the metric tensor up to a conformal factor is preserved) and affine vectors (preserving geodesics as well as affine parameter). Among these symmetries, considerable interest is shown in conformal symmetry. Along the null geodesics, conformal symmetry produces constant of motion for massless particles. Apart from their classification by such symmetries, the solutions of EFEs are also categorized according to Petrov types of the curvature and Segree types of the energy-momentum tensor. A detailed study of such a work is found in [1]. It is important to note that the Lorentzian metrics with the plane symmetry can be of Petrov type D or O [1].

In the past, the conformal symmetry was considered merely a mathematical tool in integrating the EFEs and its physical applications in cosmology and astrophysics remained unnoticed [4]. Recently, some work has appeared that explores the use of conformal symmetry in astrophysics and cosmology. Chrobok et al. [5] made an assumption for temperature vector to be a CKV in the theory of irreversible thermodynamical processes and produced some interesting results. Bohmer et al. [6] verified that the conformal factor for conformally symmetric spacetime with non-static vector fields can be interpreted in terms of tangential velocity of the test particle moving in a circular orbit. Using an assumption of spherical symmetry that admits one-parameter group of conformal vector, Mak et al. found an exact solution describing the interior of a charged strange quark star [7]. Also Usmani and others, proposed an astrophysical model, known as gravastar that admits a CKV [8]. In short, conformal symmetry has wide ranging applications, in understanding both physical as well as geometrical properties of spacetime physics.

Whereas spacetime symmetries are mostly used in understanding the physics of Lorentzian geometries, a considerable attention has also been given to studying Killing, homothetic and conformal symmetries in ultra-hyperbolic geometry. In a recent paper [9], all Killing's symmetries in complex HH-spaces with cosmological constant $\Lambda$ are found. The explicit complex metrics admitting null Killing vectors are investigated along with some Lorentzian and ultra-hyperbolic slices of these metrics [10]. Chudecki studied the conformal Killing's equations and their integrability conditions for expanding hyper-heavenly spaces with $\Lambda$ in spinorial formalism [11]. Chudecki et al. also provide a detailed study of proper CKVs in self-dual Einstein spaces [12].

As discussed above, the CKVs are motions along which spacetime metric remains unchanged up to a scale factor. A CKV is considered as a global smooth vector field $V$ over a manifold $W$, such that for smooth conformal function $\psi : W \to \Re$ of $V$, the relation $V_{e;d} = \psi g_{ed} + N_{ed}$ holds, where $g_{ed}$ are metric tensor components and $N_{ed} = (-N_{de})$ is the bivector of $V$. Mathematically, this relation is given by:

$$\mathop{L}_{V} g_{ed} = 2\psi g_{ed}, \qquad (1.1)$$

where $L_V$ represents Lie derivative along the vector field $V$ and $\psi$ depends on the chosen coordinate system. In a coordinate frame, Eq. (1.1) takes the simple form,

$$g_{ed,s} V^s + g_{ds} V^s{,}_e + g_{es} V^s{,}_d = 2\psi\, g_{ed}, \qquad (1.2)$$

where a comma in subscript represents partial derivative. If $\psi$ becomes constant, the vector $V$ reduces to a HV and in case $\psi$ vanishes, the vector becomes KV. The maximum number of CKVs for conformally flat spacetimes are discussed by Eisenhart [13] and Schouten [14]. Hall et al. published a remarkable paper [15], proving that the maximum dimension of CKV algebra in non-conformally flat spacetime is seven, while for conformally flat spacetimes it is fifteen. Kramer et al. considered certain assumptions (in relation with Lie algebra) in terms of rigidly rotating stationary axisymmetric perfect fluid spacetimes admitting CKVs and concluded that under such restrictions EFEs do not have a solution [16]. Maartens et al. obtained CKVs for both conformally flat and non-conformally flat static spherically symmetric spacetimes [17]. These results show that when the spacetime is non-conformally flat, it admits at most two proper CKVs. In addition, they concluded that eleven proper CKVs exist when the spacetime is conformally flat. Shabbir et al., investigated CKVs admitted by Bianchi type VIII and IX and spatially homogeneous rotating spacetimes [18, 19]. They showed that spatially homogenous rotating spacetimes do not admit proper CKVs. Considering Robertson-Walker spacetime; Martens et al. obtained CKVs that are neither normal nor tangent to spacelike homogeneous hypersurfaces [20]. Moopanar and Maharaj investigated CKVs in non-static spherically symmetric spacetimes [21]. Considering shear free spherically symmetric spacetimes, Moopanar et al. gave a complete discussion of conformal geometry without specifying the matter content [22]. Furthermore, Saifullah et al. investigated conformal motions of plane symmetric static spacetimes and found that no proper CKV exist when the plane symmetric static spacetime is non-conformally flat [23]. More recently, CKVs are also studied for different spacetime metrics in teleparallel theory of gravitation, such as static cylindrically symmetric, Bianchi type I and static plane symmetric spacetimes [24-26].

In general relativity theory, CKVs have wide range of applications. They play their role, not only at the geometric level, but also at the kinematics and dynamics levels [27]. Kinematic variables such as rotation, expansion and shear can be studied under the assumption that a spacetime admit

CKVs. The CKVs help us in the investigation of such variables by imposing certain restrictions on them. These variables are then used to produce well known results, some of which can be seen in [17, 28, 29]. Conformal Killing vectors have an important role at dynamics level as well. Some physically plausible solutions of Einstein's field equations have been obtained in [30-33] under the assumption that the spacetime admit CKVs. At the geometric level it is well understood that application of CKVs makes possible coordinate choice to simplify the metric. This fact can be seen in [2, 34]. The wide range applications of CKVs in astrophysics and cosmology (as discussed above) and at geometric, dynamics and kinematic levels motivated us to explore the CKVs of plane symmetric non static spacetimes. The results obtained in this paper can further be applied to study the dynamical and kinematic properties of the spacetime under consideration.

This paper is organized as follows: In section 2, we write ten conformal Killing's equations in a plane symmetric spacetime and derive a general form of CKV components and the conformal factor $\psi$. In section 3, we list conditions that are imposed on the form of CKVs. These particular forms include timelike and inheriting CKVs. Section 4 is devoted to obtaining CKVs for certain plane symmetric spacetime metrics. In particular, section 4(a) gives the method in which CKVs are obtained in some special non-conformally flat plane symmetric spacetimes, while in section 4(b), we show how CKVs are obtained in some conformally flat plane symmetric spacetimes. A brief summary and discussion of the results is given in the last section.

## 2. GENERAL FORM OF CONFORMAL KILLING EQUATIONS AND CONFORMAL KILLING VECTOR COMPONENTS

We take the most general line element of plane symmetric spacetimes in its usual $(t,x,y,z)$ coordinates as [1],

$$ds^2 = -e^{2A(t,x)}dt^2 + e^{2C(t,x)}dx^2 + e^{2B(t,x)}[dy^2 + dz^2], \tag{2.1}$$

where $A$, $B$ and $C$ are functions of t and x only. The metric (2.1) admits a minimal set of three independent spatial KVs given by $\partial_y$, $\partial_z$ and $z\partial_y - y\partial_z$. In this minimal set the first two KVs represent conservation of linear momentum along 'y' and 'z' directions, whereas the third represents conservation of angular momentum. In case the above metric becomes static, it admits an additional timelike KV $\partial_t$. Expanding Eq. (1.2) with the help of Eq. (2.1), we have the following system of ten coupled partial differential equations,

$$A_t(t,x)V^0 + A_x(t,x)V^1 + V^0_{,0} = \psi(t,x,y,z), \tag{2.2}$$

$$e^{2C(t,x)}V^1_{,0} - e^{2A(t,x)}V^0_{,1} = 0, \tag{2.3}$$

$$e^{2B(t,x)}V^2_{,0} - e^{2A(t,x)}V^0_{,2} = 0, \tag{2.4}$$

$$e^{2B(t,x)}V^3_{,0} - e^{2A(t,x)}V^0_{,3} = 0, \tag{2.5}$$

$$C_t(t,x)V^0 + C_x(t,x)V^1 + V^1_{,1} = \psi(t,x,y,z), \tag{2.6}$$

$$e^{2B(t,x)}V^2_{,1} + e^{2C(t,x)}V^1_{,2} = 0, \tag{2.7}$$

$$e^{2B(t,x)}V^3_{,1} + e^{2C(t,x)}V^1_{,3} = 0, \tag{2.8}$$

$$B_t(t,x)V^0 + B_x(t,x)V^1 + V^2_{,2} = \psi(t,x,y,z), \tag{2.9}$$

$$V^2_{,3} + V^3_{,2} = 0, \tag{2.10}$$

$$B_t(t,x)V^0 + B_x(t,x)V^1 + V^3_{,3} = \psi(t,x,y,z). \tag{2.11}$$

To solve the above system of ten coupled equations, we first use some equations to obtain the components $V^0$, $V^1$, $V^2$ and $V^3$ of the CKV and the conformal function $\psi(t,x,y,z)$. The whole process is described briefly in the following:

Differentiating Eqs. (2.4) and (2.7) with respect to $z$, Eqs. (2.5) and (2.8) with respect to $y$ and Eq. (2.10) with respect to $t$ and $x$, we find that $V^2_{,03} = V^0_{,23} = V^1_{,23} = V^2_{,13} = 0$. Also, subtracting Eq. (2.6) from Eq. (2.9) and differentiating the resulting equation with respect to $y$ and $z$ and then using the result in Eq. (2.10), we obtain,

$$V^2 = y\left\{\frac{z^2}{2}G^1(x) + zG^2(x)\right\} + \frac{z^2}{2}G^3(x) + zG^4(x) - \frac{y^3}{6}G^1(x) - \frac{y^2}{2}G^3(x) + yP^1(t,x) + P^3(t,x),$$

$$V^3 = -\frac{y^2}{2}\left\{zG^1(x) + G^2(x)\right\} - y\left\{zG^3(x) + G^4(x)\right\} + \frac{z^3}{6}G^1(x) + \frac{z^2}{2}G^2(x) + zP^1(t,x) + P^2(t,x),$$

where $G^i(x)$ and $P^k(t,x)$, $i=1,2,3,4$ and $k=1,2,3$ are functions of integration. Using above expressions in Eqs. (2.4), (2.5), (2.7), (2.8) and relations $V^2_{,03} = V^0_{,23} = V^1_{,23} = V^2_{,13} = 0$ simultaneously, we obtain the following form of CKV components and the conformal factor $\psi$:

$$V^0 = e^{2(B-A)}\left\{\frac{z^2}{2}P^1_t(t,x) + zP^2_t(t,x)\right\} + e^{2(B-A)}\left\{\frac{y^2}{2}P^1_t(t,x) + yP^3_t(t,x)\right\} + P^0(t,x),$$

$$V^1 = -e^{2(B-C)}\left\{\frac{z^2}{2}P^1_x(t,x) + zP^2_x(t,x)\right\} - e^{2(B-C)}\left\{\frac{y^2}{2}P^1_x(t,x) + yP^3_x(t,x)\right\} + P^4(t,x),$$

$$V^2 = d_1 zy + \frac{d_2}{2}z^2 + c_3 z - \frac{d_2}{2}y^2 + yP^1(t,x) + P^3(t,x),$$

$$V^3 = -d_2 zy + \frac{d_1}{2}z^2 - c_3 y - \frac{d_1}{2}y^2 + zP^1(t,x) + P^2(t,x),$$

$$\psi(t,x,y,z) = C_t e^{2(B-A)}\left\{\frac{z^2}{2}P^1_t(t,x) + zP^2_t(t,x)\right\} + C_t e^{2(B-A)}\left\{\frac{y^2}{2}P^1_t(t,x) + yP^3_t(t,x)\right\} - e^{2(B-C)}\left\{\frac{z^2}{2}P^1_{xx}(t,x) + zP^2_{xx}(t,x)\right\}$$

$$-e^{2(B-C)}\left\{\frac{y^2}{2}P^1_{xx}(t,x) + yP^3_{xx}(t,x)\right\} + (C_x - 2B_x)e^{2(B-C)}\left\{\frac{z^2}{2}P^1_x(t,x) + zP^2_x(t,x)\right\} + C_t P^0(t,x)$$

$$+ (C_x - 2B_x)e^{2(B-C)}\left\{\frac{y^2}{2}P^1_x(t,x) + yP^3_x(t,x)\right\} + C_x P^4(t,x) + P^4_x(t,x),$$

where $d_1, d_2, c_3 \in \Re$ and the functions $P^k(t,x)$, $k=0,1,2,3,4$ and their derivatives arise in the process of integration and need to be determined. The above CKV components and the conformal factor are subject to the following integrability conditions:

$$P^1_{tx}(t,x) + \{B_t - C_t\}P^1_x(t,x) + \{B_x - A_x\}P^1_t(t,x) = 0, \tag{2.12}$$

$$P^2_{tx}(t,x) + \{B_t - C_t\}P^2_x(t,x) + \{B_x - A_x\}P^2_t(t,x) = 0, \tag{2.13}$$

$$P^3_{tx}(t,x) + \{B_t - C_t\}P^3_x(t,x) + \{B_x - A_x\}P^3_t(t,x) = 0, \tag{2.14}$$

$$\{2B_x - A_x - C_x\}e^{-2C}P^1_x(t,x) + e^{-2A}P^1_{tt}(t,x) + e^{-2C}P^1_{xx}(t,x) + \{2B_t - A_t - C_t\}e^{-2A}P^1_t(t,x) = 0, \tag{2.15}$$

$$\{2B_x - A_x - C_x\}e^{-2C}P^2_x(t,x) + e^{-2A}P^2_{tt}(t,x) + e^{-2C}P^2_{xx}(t,x) + \{2B_t - A_t - C_t\}e^{-2A}P^2_t(t,x) = 0, \tag{2.16}$$

$$\{2B_x - A_x - C_x\}e^{-2C}P^3_x(t,x) + e^{-2A}P^3_{tt}(t,x) + e^{-2C}P^3_{xx}(t,x) + \{2B_t - A_t - C_t\}e^{-2A}P^3_t(t,x) = 0, \tag{2.17}$$

$$\{B_t - C_t\}e^{-2A}P^1_t(t,x) + \{B_x - C_x\}e^{-2C}P^1_x(t,x) + e^{-2C}P^1_{xx}(t,x) = 0, \tag{2.18}$$

$$\{B_t - C_t\}e^{-2A}P^2{}_t(t,x) + \{B_x - C_x\}e^{-2C}P^2{}_x(t,x) + e^{-2C}P^2{}_{xx}(t,x) = -d_1 e^{-2B}, \tag{2.19}$$

$$\{B_t - C_t\}e^{-2A}P^3{}_t(t,x) + \{B_x - C_x\}e^{-2C}P^3{}_x(t,x) + e^{-2C}P^3{}_{xx}(t,x) = d_2 e^{-2B}, \tag{2.20}$$

$$e^{2A}P^0{}_x(t,x) - e^{2C}P^4{}_t(t,x) = 0, \tag{2.21}$$

$$\{A_t - C_t\}P^0(t,x) + \{A_x - C_x\}P^4(t,x) + P^0{}_t(t,x) - P^4{}_x(t,x) = 0, \tag{2.22}$$

$$\{B_t - C_t\}P^0(t,x) + \{B_x - C_x\}P^4(t,x) + P^1(t,x) - P^4{}_x(t,x) = 0. \tag{2.23}$$

At this stage, we introduce new variables $\beta_i = (\beta_1, \beta_2, \beta_3) = \left(\dfrac{y^2 + z^2}{2}, z, y\right)$ and $P^i = (P^1, P^2, P^3)$. In these variables, the components of the CKV and the conformal factor can be rewritten in a more convenient form using the Einstein's summation convention as,

$$V^0 = e^{2(B-A)}\beta_i P^i_t + P^0, \qquad V^1 = -e^{2(B-C)}\beta_i P^i_x + P^4,$$

$$V^2 = (\beta_i)_{,2} P^i + d_1 zy + \frac{d_2}{2}(z^2 - y^2) + c_3 z, \qquad V^3 = (\beta_i)_{,3} P^i - d_2 zy + \frac{d_1}{2}(z^2 - y^2) - c_3 y,$$

$$\psi(t,x,y,z) = C_t e^{2(B-A)}\beta_i P^i_t - e^{2(B-C)}\beta_i P^i_{xx} + (C_x - 2B_x)e^{2(B-C)}\beta_i P^i_x + C_t P^0 + C_x P^4 + P^4_x.$$

Like-wise, the twelve integrability conditions given by Eqs. (2.12)-(2.23) reduce to,

$$P^i{}_{tx} + \{B_t - C_t\}P^i{}_x + \{B_x - A_x\}P^i{}_t = 0, \tag{2.24}$$

$$(2B_x - A_x - C_x)e^{-2C}P^i{}_x + e^{-2A}P^i{}_{tt} + e^{-2C}P^i{}_{xx} + (2B_t - A_t - C_t)e^{-2A}P^i{}_t = 0, \tag{2.25}$$

$$(B_t - C_t)e^{2(B-A)}P^i{}_t + (B_x - C_x)e^{2(B-C)}P^i{}_x + e^{2(B-C)}P^i{}_{xx} = k_i, \tag{2.26}$$

$$P^0_x - e^{2(C-A)}P^4_t = 0, \tag{2.27}$$

$$(A_t - C_t)P^0 + (A_x - C_x)P^4 + P^0{}_t - P^4{}_x = 0, \tag{2.28}$$

$$(B_t - C_t)P^0 + (B_x - C_x)P^4 + P^1 - P^4{}_x = 0, \tag{2.29}$$

where $k_i = 0, -d_1, d_2$, for $i = 1, 2, 3$ respectively. In order to solve the above integrability conditions completely, we impose certain restrictions either on the metric functions or the CKV components. To this end, we restrict the components of the CKVs to admit a particular form and present results in the next section.

## 3. CONFORMAL KILLING VECTORS OF PARTICULAR FORMS

(I) In this section, we discuss some cases in which CKVs admit a particular form in which the

timelike CKV is orthogonal to the orbits of the planar symmetry, i.e. with only t and x components. Assuming a suitable coordinate transformation, which preserves the form of metric given by Eq. (2.1), allows one to assume that the CKV has only a temporal component. In this case, the CKV becomes a purely timelike vector, i.e. $V = (V^0, 0, 0, 0)$. For consistency, we must have $P^i = P^4 = d_1 = d_2 = c_3 = 0$ and $V^0 = P^0$. The integrability condition given by Eq. (2.27) then implies that $V^0 = P^0(t)$. Since $P^0 \neq 0$, Eqs. (2.29) and (2.28) respectively give $B_t = C_t$ and $(A_t - C_t)P^0 + P^0_t = 0$. Integrating the later equation instantly yields $P^0 \propto e^{C-A}$, indicating existence of a timelike CKV parallel to the time like vector $u^a$, defined as $u^a = e^{-A}\delta^a_0$.

(II) In their paper, Herrera et al. [35] introduced a condition $L_V u_a = \psi u_a$, where $u^a$ is taken as four velocity of the co-moving fluid. This condition is called the inheriting condition. Here, we use this condition for a timelike vector $u^a = e^{-A}\delta^a_0$. The above mentioned inheriting condition $L_V u_a = \psi u_a$, in expanded form can be written as:

$$u_{a,b} V^b + u_b V^b_{,a} = \psi u_a . \tag{3.1}$$

Solving Eq. (3.1), it is easily found that $V^0_{,i} = 0$ for $i = 1, 2, 3$, whereas $V^0 = V^0(t)$ and the corresponding conformal factor takes the form $\psi = A_t V^0 + A_x V^1 + V^0_{,0}$. This suggests that $P^0$ depends on $t$ only, with $P^i_t = 0$ for $i = 1, 2, 3$. Also, whereas the integrability condition given by Eq. (2.27) suggests that $P^4_t = 0$, the remaining integrability conditions take the form:

$$\{B_t - C_t\} P^i{}_x = 0, \tag{3.2}$$

$$(2B_x - A_x - C_x) P^i{}_x + P^i{}_{xx} = 0, \tag{3.3}$$

$$(B_x - C_x) e^{2(B-C)} P^i{}_x + e^{2(B-C)} P^i{}_{xx} = k_i, \tag{3.4}$$

$$(A_t - C_t) P^0 + (A_x - C_x) P^4 + P^0{}_t - P^4{}_x = 0, \tag{3.5}$$

$$(B_t - C_t) P^0 + (B_x - C_x) P^4 + P^1 - P^4{}_x = 0. \tag{3.6}$$

From Eq. (3.2) it is easy to note that two possibilities arise, namely, $B_t - C_t \neq 0$ and $B_t - C_t = 0$. A complete solution of the integrability conditions is found in the first case, while in the second case the solutions are arbitrary and will not be presented.

Using $B_t - C_t \neq 0$ in Eq. (3.2), it is immediately noticed that $P^i{}_x = 0$. Using this fact in Eq. (3.4), one find that $k_i = 0$. Now subtracting the remaining Eqs. (3.5) and (3.6) give:

$$(B_t - A_t)P^0 + (B_x - A_x)P^4 + P^1 - P^0_t = 0. \tag{3.7}$$

After some manipulations, it is easily found that the CKVs and their corresponding conformal factor take the form,

$$V^0 = P^0, V^1 = P^4, V^2 = (\beta_i)_{,2} P^i + c_3 z, V^3 = (\beta_i)_{,3} P^i - c_3 y,$$

$$\psi(t, x, y, z) = C_t P^0 + C_x P^4 + P^4_x,$$

subject to a solution of equation (3.7). Here it is worth noting that one CKV is the usual KV giving rotational symmetry $z\partial_y - y\partial_z$, while all other CKVs are arbitrary functions of t and x.

## 4. CONFORMAL KILLING VECTORS OF SOME SPECIAL PLANE SYMMETRIC SPACETIME METRICS

In this section, we investigate certain CKVs for some special classes of plane symmetric spacetime metrics. These spacetime metrics are chosen from the literature and are obtained either by solving the EFEs under certain assumptions or by imposing some symmetry restrictions on the spacetime metric. It is worth mentioning here that throughout this section, the constants $c_i$ are numbered so that $c_1, c_2, c_3$ represent the three spatial KVs $\partial_y$, $\partial_z$ and $z\partial_y - y\partial_z$ respectively, representing two linear momentum (along y and z) and one angular momentum conservations (along x ). Also in the static case, $c_0$ is chosen to correspond with the timelike KV $\partial_t$, giving energy conservation.

### 4(a) NON-CONFORMALLY FLAT PLANE SYMMETRIC SPACETIMES AND THEIR CONFORMAL KILLING VECTORS

**Case (I):** To solve the integrability conditions found above completely, we consider a plane symmetric non-conformally flat metric [36],

$$ds^2 = -x^{2+2\sqrt{2}} dt^2 + dx^2 + x^2(dy^2 + dz^2) \tag{4.1}$$

The above metric is static and admits at least four independent KVs $\partial_t$, $\partial_y$, $\partial_z$ and $z\partial_y - y\partial_z$. The minimal set of KVs represents existence of energy conservation, two linear momentum conservations (along y and z) and angular momentum conservation. The non-zero anisotropic energy-momentum tensor components for the metric represented by Eq. (4.1) are given by $T_{00} = \frac{1}{4}x^{\sqrt{2}-1}$, $T_{11} = \frac{3+2\sqrt{2}}{x^2}$, $T_{22} = T_{33} = \frac{\sqrt{2}+1}{4}x^{-2(\sqrt{2}+1)}$. The CKVs and the associated conformal factor admitted by the above metric are given by,

$$V^0 = -\frac{c_4}{2\sqrt{2}}x^{-2\sqrt{2}} - \frac{c_4}{\sqrt{2}}t^2 - \sqrt{2}c_5 t + c_0, \qquad V^1 = c_4 xt + c_5 x$$

$$V^2 = c_3 z + c_1, \qquad\qquad\qquad\qquad\qquad V^3 = -c_3 y + c_2,$$

$$\psi(t,x,y,z) = c_4 t + c_5,$$

where $c_0, c_1, c_2, c_3, c_4, c_5 \in \mathfrak{R}$. This result reveals that the above plane symmetric metric given by Eq. (4.1) admits six independent CKVs of which one is proper CKV given by $-\left(\frac{1}{2\sqrt{2}}x^{-2\sqrt{2}} + \frac{1}{\sqrt{2}}t^2\right)\partial_t + xt\partial_x$. Also the dimension of the homothetic symmetry group is five with one proper HV given by $-\sqrt{2}t\partial_t + x\partial_x$. Note also that the dimension of the isometry group is four. Now a CKV is called special CKV if $\psi_{;ab} = 0$ [3]. This suggests that the above CKVs are special CKVs. A six dimensional group of CKVs give the generators of the conformal symmetry group:

$$X_0 = \partial_t, \quad X_1 = \partial_y, \quad X_2 = \partial_z, \quad X_3 = z\partial_y - y\partial_z, \quad X_4 = -\left(\frac{1}{2\sqrt{2}}x^{-2\sqrt{2}} + \frac{1}{\sqrt{2}}t^2\right)\partial_t + xt\partial_x,$$

$$X_5 = -\sqrt{2}t\partial_t + x\partial_x,$$

**Case (II):** Another plane symmetric non-conformally flat metric, admitting self-similarity of second kind [36], is given by,

$$ds^2 = -dt^2 + dx^2 + 2t\left(dy^2 + dz^2\right) \tag{4.2}$$

This is a non-static metric and admits three independent KVs. The an-isotropic non-zero energy-momentum tensor components for this metric are $T_{00} = \frac{1}{16t^2}$, $T_{11} = \frac{5}{16t^2}$, $T_{22} = T_{33} = \frac{3\sqrt{2}}{16}t^{\frac{-3}{2}}$. As for the integrability conditions are concerned, they are completely solved and the CKVs admitted by this metric, along with the conformal factor, are given by,

$$V^0 = 2c_4 t, \quad V^1 = 2c_4 x + c_5, \quad V^2 = c_3 z + c_4 y + c_1, \quad V^3 = -c_3 y + c_4 z + c_2,$$
$$\psi(t, x, y, z) = 2c_4,$$

where $c_1, c_2, c_3, c_4, c_5 \in \Re$. Thus the metric given by Eq. (4.2) admits five independent CKVs with no proper CKV. In fact, the CKVs for the metric represented by Eq. (4.2) are just HVs, where the dimension of homothetic group is five, with one proper HV given by $t\partial_t + x\partial_x + \frac{y}{2}\partial_y + \frac{z}{2}\partial_z$. Also, the above metric admits four independent KVs given by $\partial_y$, $\partial_z$, $z\partial_y - y\partial_z$ and $\partial_x$. The generators of the five dimensional group of CKVs are:

$$X_1 = \partial_y, \quad X_2 = \partial_z, \quad X_3 = z\partial_y - y\partial_z, \quad X_4 = t\partial_t + x\partial_x + \frac{1}{2}y\partial_y + \frac{1}{2}z\partial_z, \quad X_5 = \partial_x.$$

The above symmetry groups corresponds to three linear momentum conservation laws given by $X_1$, $X_2$, and $X_5$, one angular momentum conservation given by $X_3$ and one scaling conservation law. Here, it is important to mention that the metric given by Eq. (4.2) can also be obtained by solving the integrability conditions with $A(t, x) = 0$, $B_x(t, x) = 0$ and $C(t, x) = 0$. In this case, the metric becomes,

$$ds^2 = -dt^2 + dx^2 + (c_6 t)^{2(\frac{c_6 - c_8}{c_6})}(dy^2 + dz^2). \tag{4.3}$$

Re-defining $c_6 = 2$ and $c_8 = 1$, one recovers the metric given by Eq. (4.2).

**Case (III):** Here, we consider plane symmetric spacetime metric admitting a five dimensional isometry group. This metric is taken from [37] and is given by,

$$ds^2 = -e^{\frac{2x}{a}} dt^2 + dx^2 + e^{\frac{2x}{b}}(dy^2 + dz^2), \quad a \neq b \neq 0. \tag{4.4}$$

For the above metric, the non-zero energy-momentum tensor components are given by,

$$T_{00} = \frac{-3}{4b^2} e^{\frac{x}{a}}, \quad T_{11} = \frac{1}{4}\left\{\frac{1}{b^2} + \frac{2}{ab}\right\}, \quad T_{22} = T_{33} = e^{\frac{x}{b} - \frac{x}{a}}\left\{\frac{1}{b^2} + \frac{1}{ab} + \frac{1}{a^2}\right\}.$$ In this case, the conformal factor, $\psi(t, x, y, z)$, becomes zero and hence the CKVs become same as the KVs given by,

$$V^0 = \frac{b}{a} c_4 t + c_0, \quad V^1 = -c_4 b, \quad V^2 = c_3 z + c_4 y + c_1, \quad V^3 = -c_3 y + c_4 z + c_2,$$

The Lie algebra structure of this example and for the coming examples can be found in [37].

**Case (IV):** In this case, we have solved the integrability conditions completely by considering the following plane symmetric spacetime metric [37]:

$$ds^2 = -e^{\frac{2x}{a}} dt^2 + dx^2 + dy^2 + dz^2, \quad a \neq 0. \tag{4.5}$$

The above metric admits five independent KVs, in which four are included in the list given in section 2. The non-zero energy-momentum tensor components for this metric are $T_{22} = T_{33} = \frac{1}{4a^2} e^{-\frac{x}{a}}$. For this metric, it is found that the conformal factor $\psi(t, x, y, z)$ is zero, and therefore the CKVs are just KVs as given in [37].

**Case (V):** In this case, we consider another particular form of plane symmetric spacetime metric given by,

$$ds^2 = -\cos^2\frac{x}{a} dt^2 + dx^2 + dy^2 + dz^2, a \neq 0. \tag{4.6}$$

This metric is also taken from the reference [37], which admits six independent KVs. The non-zero energy-momentum tensor components for this metric take the form, $T_{22} = T_{33} = -\frac{1}{4a^2}\sec\frac{x}{a}\left\{1 + \sec^2\frac{x}{a}\right\}$. Using Eq. (4.6), one can easily solve the integrability conditions completely to find that it does not admit proper CKV. We have also solved the integrability conditions completely for the following two plane symmetric spacetime metrics taken from [37]

$$ds^2 = -dt^2 + \cos^2\frac{t}{a} dx^2 + dy^2 + dz^2, \quad a \neq 0. \tag{4.7}$$

$$ds^2 = -dt^2 + e^{\frac{2t}{a}} dx^2 + dy^2 + dz^2, \quad a \neq 0. \tag{4.8}$$

Both metrics admit six independent KVs. On solving the integrability conditions separately for the above metrics, it is easily found that the CKVs are same as the KVs as obtained in [37].

**Case (VI):** In this case, we assume that $A = 0$ and $B$ and $C$ are functions of $t$ coordinate only. In the light of these assumptions, the spacetime metric given by Eq. (2.1) takes the form,

$$ds^2 = -dt^2 + e^{2C(t)} dx^2 + e^{2B(t)}[dy^2 + dz^2], \tag{4.9}$$

For the above metric, the non-vanishing Ricci tensor components are given by,

$R_{00} = -[2B''(t) + C''(t) + 2B'^2(t) + C'^2(t)]$, $\quad R_{11} = e^{2C(t)}[C'^2(t) + 2B'(t)C'(t) + C''(t)]$ and $R_{22} = R_{33} = e^{2B(t)}[2B'^2(t) + B'(t)C'(t) + B''(t)]$. In order to obtain a vacuum solution, we require that all the Ricci tensor components are zero, i.e., $R_{00} = R_{11} = R_{22} = R_{33} = 0$. Solving these three equations simultaneously, we can easily find that $B(t) = \ln\left\{\frac{3}{2}(mt + c_4)\right\}^{\frac{2}{3}}$ and $C(t) = \ln(mt + c_4)^{-\frac{1}{3}}$, where $m$ and $c_4$ are constants. The constant $c_4$ can be removed by an obvious linear change of the $t$ coordinate. Under this transformation, the metric functions can be rewritten as $B(t) = \ln\left\{\frac{3}{2}(mt)\right\}^{\frac{2}{3}}$ and $C(t) = \ln(mt)^{-\frac{1}{3}}$. Using these values of $B(t)$ and $C(t)$, we can completely solve the integrability conditions to find that the conformal factor is constant. Thus, the components of the CKV in this case take the form:

$$V^0 = 3c_5 t, \quad V^1 = 3c_5 x + c_6, \quad V^2 = c_3 z + c_5 y + c_1, \quad V^3 = -c_3 y + c_5 z + c_2,$$
$$\psi = 3c_5.$$

From above, it can be easily seen that there are five independent CKVs with no proper CKV. Also, the dimension of the homothetic symmetry group is five, with one proper HV and four KVs. In generator form, the proper HV (choosing $c_5 = 1$) is written as $3t\,\partial_t + 3x\,\partial_x + y\,\partial_y + z\,\partial_z$.

## 4(b) CONFORMALLY FLAT PLANE SYMMETRIC SPACETIMES AND THEIR CONFORMAL KILLING VECTORS

In this section, we investigate CKVs of some conformally flat classes of plane symmetric spacetimes. These conformally flat plane symmetric spacetimes are obtained by imposing certain conditions on the metric functions or they have been taken among a wide class of known conformally flat plane symmetric spacetimes from [37]. It is important to note that, any conformally flat plane symmetric spacetime will always admit a fifteen dimensional algebra of CKVs. We show that various integrability conditions given in (2.24)-(2.29) can be solved completely for some known conformally flat plane symmetric metrics obtaining fifteen CKVs in each case.

**Case (VII):** In this case, we consider the conformally flat plane symmetric spacetime whose line element is given by [37],

$$ds^2 = -dt^2 + dx^2 + e^{\frac{2x}{a}}(dy^2 + dz^2), \qquad a \neq 0. \tag{4.10}$$

In [37], it is shown that the above metric admits seven independent KVs. Solving the integrability conditions given in Eqs. (2.24)-(2.29), it is easy to find that the above metric admits a fifteen dimensional algebra of CKVs given by,

$$V^0 = \frac{1}{2a}(y^2+z^2)e^{\frac{x}{a}}\left(-c_4 e^{-\frac{t}{a}} + c_5 e^{\frac{t}{a}}\right) + \frac{z}{a}e^{\frac{x}{a}}\left(-c_6 e^{-\frac{t}{a}} + c_7 e^{\frac{t}{a}}\right) + \frac{y}{a}e^{\frac{x}{a}}\left(-c_8 e^{-\frac{t}{a}} + c_9 e^{\frac{t}{a}}\right)$$
$$- \frac{a}{2}e^{-\frac{x}{a}}\left(c_4 e^{-\frac{t}{a}} - c_5 e^{\frac{t}{a}}\right) - e^{\frac{x}{a}}\left(c_{10} e^{-\frac{t}{a}} - c_{11} e^{\frac{t}{a}}\right) + c_0,$$

$$V^1 = \frac{1}{2a}(y^2+z^2)e^{\frac{x}{a}}\left(c_4 e^{-\frac{t}{a}} + c_5 e^{\frac{t}{a}}\right) + \frac{z}{a}e^{\frac{x}{a}}\left(c_6 e^{-\frac{t}{a}} + c_7 e^{\frac{t}{a}}\right) + \frac{y}{a}e^{\frac{x}{a}}\left(c_8 e^{-\frac{t}{a}} + c_9 e^{\frac{t}{a}}\right)$$
$$- \frac{a}{2}e^{-\frac{x}{a}}\left(c_4 e^{-\frac{t}{a}} + c_5 e^{\frac{t}{a}}\right) + e^{\frac{x}{a}}\left(c_{10} e^{-\frac{t}{a}} + c_{11} e^{\frac{t}{a}}\right) - c_{12}az + c_{13}ay - c_{14}a,$$

$$V^2 = \frac{c_{13}}{2}\left(z^2 - y^2 + a^2 e^{-\frac{2x}{a}}\right) + ye^{-\frac{x}{a}}\left(c_4 e^{-\frac{t}{a}} + c_5 e^{\frac{t}{a}}\right) + e^{-\frac{x}{a}}\left(c_8 e^{-\frac{t}{a}} + c_9 e^{\frac{t}{a}}\right) + c_{12}yz + c_{14}y + c_3 z + c_1,$$

$$V^3 = \frac{c_{12}}{2}\left(z^2 - y^2 - a^2 e^{-\frac{2x}{a}}\right) + z e^{-\frac{x}{a}}\left(c_4 e^{-\frac{t}{a}} + c_5 e^{\frac{t}{a}}\right) + e^{-\frac{x}{a}}\left(c_6 e^{-\frac{t}{a}} + c_7 e^{\frac{t}{a}}\right) - c_{13} yz - c_3 y + c_{14} z + c_2,$$

$$\psi(t,x,y,z) = \frac{1}{2a^2}(y^2+z^2)e^{\frac{x}{a}}\left(c_4 e^{-\frac{t}{a}} + c_5 e^{\frac{t}{a}}\right) + \frac{z}{a^2}e^{\frac{x}{a}}\left(c_6 e^{-\frac{t}{a}} + c_7 e^{\frac{t}{a}}\right) + \frac{y}{a^2}e^{\frac{x}{a}}\left(c_8 e^{-\frac{t}{a}} + c_9 e^{\frac{t}{a}}\right)$$

$$+ \frac{1}{2}e^{-\frac{x}{a}}\left(c_4 e^{-\frac{t}{a}} + c_5 e^{\frac{t}{a}}\right) + \frac{1}{a}e^{\frac{x}{a}}\left(c_{10} e^{-\frac{t}{a}} + c_{11} e^{\frac{t}{a}}\right)$$

For the metric given by Eq. (4.10), the non-vanishing energy-momentum tensor components take the form,

$$T_{00} = \frac{-3}{4a^2}, \quad T_{11} = \frac{1}{4a^2}, \quad T_{22} = T_{33} = \frac{1}{4a^2} e^{\frac{x}{a}}.$$

From above, it can be easily noticed that out of the fifteen CKVs, seven are KVs as obtained in [37], while eight are proper CKVs with no proper HV

**Case (VIII):** In this case, we consider another conformally flat plane symmetric spacetime whose metric is given by [37],

$$ds^2 = -dt^2 + dx^2 + e^{\frac{2t}{a}}\left(dy^2 + dz^2\right), \quad a \neq 0. \tag{4.11}$$

In [37], it is shown that the above metric also admits seven independent KVs. The non-zero energy-momentum tensor components for the above spacetime metric are given by $T_{00} = \frac{1}{4a^2}, T_{11} = \frac{-3}{4a^2}, T_{22} = T_{33} = -\frac{1}{4a^2} e^{\frac{t}{a}}$. On solving the integrability conditions given by Eqs. (2.24)-(2.29) for the above metric, we see that it also admits fifteen CKVs of which seven are KVs [37] and eight are proper CKVs with no proper HV. The CKVs and the conformal factor obtained in this case are same as in case (VII) with the only difference that the variables $x$ and $t$ are interchanged.

**Case (IX):** In order to deal with this case, we impose certain restrictions on $A$ and $B$ as functions of $x$ with $A = B$ and $C = 0$. In the light of these choices, we obtain a conformally flat plane symmetric spacetime whose metric is given by,

$$ds^2 = dx^2 + e^{2A(x)}[-dt^2 + dy^2 + dz^2].  \qquad (4.12)$$

For this metric, all the Weyl tensor components vanish, whereas the non-zero energy-momentum tensor components take the form,

$$T_{00} = \frac{-1}{4} e^{A(x)} \{4A_{xx}(x) + 3A_x^2(x)\}, T_{11} = \frac{3}{4} A_x^2(x), \ T_{22} = T_{33} = \frac{1}{4} \{4A_{xx}(x) + 3A_x^2(x)\}.$$

In this case, our results are special case of those obtained for Friedmann-Robertson-Walker metric [20] with $x$ and $t$ interchange. The CKVs along with the conformal factor in this case take the form:

$$V^0 = \frac{c_4}{2}(t^2 + y^2 + z^2) + c_5 tz - c_6 ty + c_7 z + c_8 y + c_9 t + (c_{10}t + c_{11})\int e^{-A(x)} dx + c_4 \int \left\{ e^{-A(x)} \int e^{-A(x)} dx \right\} dx + c_0$$

$$V^1 = -\frac{c_{10}}{2} e^{A(x)}(-t^2 + y^2 + z^2) + c_5 z e^{A(x)} \int e^{-A(x)} dx - c_{13} z e^{A(x)} - c_{10} y e^{A(x)} \int e^{-A(x)} dx - c_{14} y e^{A(x)}$$

$$+ c_4 t e^{A(x)} \int e^{-A(x)} dx + c_9 e^{A(x)} \int e^{-A(x)} dx + c_{10} e^{A(x)} \int \left\{ e^{-A(x)} \int e^{-A(x)} dx \right\} dx + e^{A(x)} \left\{ \frac{c_{10}}{2} t^2 + c_{11} t + c_{12} \right\}$$

$$V^2 = \frac{c_6}{2}(-t^2 - y^2 + z^2) + c_5 yz + c_4 ty + c_3 z + c_9 y + c_8 t + (c_{10} y + c_{14})\int e^{-A(x)} dx + c_6 \int \left\{ e^{-A(x)} \int e^{-A(x)} dx \right\} dx + c_1$$

$$V^3 = -\frac{c_5}{2}(-t^2 + y^2 - z^2) - c_6 yz + c_4 tz + c_9 z - c_3 y + c_7 t + (c_4 z + c_{13})\int e^{-A(x)} dx - c_5 \int \left\{ e^{-A(x)} \int e^{-A(x)} dx \right\} dx + c_2$$

$$\psi(t, x, y, z) = -\frac{c_{10}}{2}(y^2 + z^2) A_x(x) e^{A(x)} + c_5 z e^{2A(x)} - c_6 y e^{2A(x)} + c_5 z A_x(x) e^{A(x)} \int e^{-A(x)} dx$$

$$- c_6 y A_x(x) e^{A(x)} \int e^{-A(x)} dx - c_{13} z A_x(x) e^{A(x)} - c_{14} y A_x(x) e^{A(x)}$$

$$+ c_4 t + c_4 t A_x(x) e^{A(x)} \int e^{-A(x)} dx + c_{10} \int e^{-A(x)} dx + c_{10} A_x(x) e^{A(x)} \int \left\{ e^{-A(x)} \int e^{-A(x)} dx \right\} dx$$

$$+ c_9 A_x(x) e^{A(x)} \int e^{-A(x)} dx + A_x(x) e^{A(x)} \left\{ \frac{c_{10}}{2} t^2 + c_{11} t + c_{12} \right\} + c_9$$

For the above metric, with arbitrary metric function $A(x)$, we obtain fifteen independent CKVs, of which eight are proper CKVs while the remaining seven are HVs. Additionally, the above metric admits one proper HV $t\partial_t + x\partial_x + y\partial_y + z\partial_z$ if all constants except $c_9$ appearing in $\psi(t, x, y, z)$ are zero and $A_x(x) e^{A(x)} \int e^{-A(x)} dx = $ constant.

## 5. SUMMARY AND DISCUSSION

Considering a general form of a plane symmetric spacetimes metric, ten conformal Killing's equations are solved and a general form of CKVs along with their conformal factor are obtained subject to twelve integrability conditions. Imposing certain conditions on components of the vector field or the metric functions, the integrability conditions are solved completely in sections 3 and 4. When subjected to one integrability condition, it is shown that purely timelike CKVs are admitted. A total of nine cases (consisting of non conformally flat and conformally flat plane symmetric spacetimes) are considered.

The physical significance of our obtained results can be established by further investigations that whether there exist any plane symmetric spacetime admitting CKVs which satisfy the energy conditions, so that they can be used as potential spacetime metrics. If such plane symmetric spacetimes exist, then it will be interesting to see if they are known solutions or they belong to a new family of solutions. After our exploration of the full conformal geometry for plane symmetric spacetimes it is now possible to examine the EFEs comprehensively and find some new solutions with conformal symmetry. In our paper we did not specify any matter distribution to find the general conformal symmetry. If one choose particular matter field, the field equations will put restrictions on dynamics. This fact can be seen in [38] where authors have shown that CKVs places particular restrictions on the dynamical behavior of the model and the gravitational field. Different matter fields are likely to produce different results, for example the presence of non zero electromagnetic field may produce new effects which are absent in resulting for neutral matter.

In this paper we did not solve the EFEs for any matter field but to analyze the effects of conformal geometry on a specific matter field we have taken some example metrics from literature, some of which were obtained by solving the EFEs. In case-I we have taken metric (4.1) form [36] and discussed its CKVs. This metric is obtained by solving EFEs and equation of motion under the assumption that the energy density $\rho$ and pressure $p$ satisfy an equation $p = \kappa \rho$ for $\kappa = -3 \pm \sqrt{2}$. Thus case-I show that under these conditions a solution of EFEs exist with six special CKVs, among which one is proper CKV. The minimal set of KVs represents existence of energy conservation, two linear momentum conservations (along y and z) and angular momentum conservation. Similarly metric (4.2) of case-II was obtained in [36] for different choices of the energy density $\rho$ and pressure $p$. In this case we showed that the spacetime metric admit no proper CKV. We obtained a plane symmetric vacuum solution of EFEs (case-VI) and it is shown that such metric do not admit proper CKV. In cases III-V, we only obtained solutions being KVs. Considering a special form of the plane symmetric spacetimeIn cases VII-IX, some special conformally flat plane symmetric spacetimes are considered, for each of which a fifteen dimensional algebra of CKVs is obtained. In cases VII and VIII, the spacetime metrics admit eight proper CKVs with no proper HV. Only in case IX, the spacetime metric admits eight proper CKVs and one proper HV for some particular choice of the metric functions.

Since, we have explored the general form of the CKV components, conformal factor and integrability conditions for plane symmetric spacetime and solved them completely for some special classes; it may be of interest to extend this analysis to spacetimes admitting plane conformal symmetry.

## ACKNOWLEDGMENTS

We would like to thank the unknown referees for their valuable suggestions and comments. In the light of their suggestions and comments, we believe that both content and presentation of the paper has significantly improved.